\documentclass[prl,twocolumn,showpacs,floatfix,superscriptaddress]{revtex4}
\usepackage{graphicx,epsfig,color,dcolumn}

\begin{document}

\title{Band Offsets at the Si/SiO$_2$ Interface from Many-Body Perturbation
Theory}

\author{R. Shaltaf}
\author{G.-M. Rignanese}
\author{X. Gonze}
\affiliation{European Theoretical Spectroscopy Facility (ETSF)
and Unit\'e Physico-Chimie et de Physique des Mat\'eriaux (PCPM),\\
Universit\'e catholique de Louvain, 1 Place Croix du Sud, B-1348
Louvain-la-Neuve Belgique}
\author{Feliciano Giustino}
\altaffiliation{Present address: Department of Physics,
University of California at Berkeley,
Berkeley, California 94720, USA}
\author{Alfredo Pasquarello}
\affiliation{Ecole Polytechnique F\'ed\'erale de Lausanne (EPFL),
Institute of Theoretical Physics, CH-1015 Lausanne, Switzerland}
\affiliation{Institut Romand de Recherche Num\'erique en Physique des
Mat\'eriaux (IRRMA), CH-1015 Lausanne, Switzerland}

\date{\today}

\begin{abstract}

We use many-body perturbation theory, the state-of-the-art method for band
gap calculations, to compute the band offsets at the Si/SiO$_2$ interface. We
examine the adequacy of the usual approximations in this context. We show
that (i) the separate treatment of band-structure and potential lineup
contributions, the latter being evaluated within density-functional theory,
is justified, (ii) most plasmon-pole models lead to inaccuracies in the
absolute quasiparticle corrections, (iii) vertex corrections can be
neglected, (iv) eigenenergy self-consistency is adequate.  Our theoretical
offsets agree with the experimental ones within 0.3 eV.

\end{abstract}


\maketitle

The band offsets (BOs) are among the most important properties of a
heterostructure.  Their precise knowledge is extremely important to engineer
electronic~\cite{ Wilk2001} and optoelectronic~\cite{ Morteani2004} devices.
Various theoretical methods have been used to estimate the BOs~\cite{
Tejedor1978-Harrison1980-Tersoff1984-Robertson2000-Peacock2002,
VandeWalle1986, Yamasaki2001}.  Among these, density functional theory (DFT)
calculations allow one to study the variation of interface dipoles with the
interface structure, without resorting to experimental input~\cite{
VandeWalle1986}.

In the DFT approach, the valence band offset (VBO) and the conduction band
offset (CBO) are conveniently split into two terms: $\text{(V,C)BO}=\Delta
E_{v,c}^{DFT} + \Delta V$.  The first term $\Delta E_{v,c}^{DFT}$, referred
to as the {\it band-structure contribution}, is defined as the difference
between the valence band maxima (VBM) or the conduction band minima (CBM)
{\it relative to the average of the electrostatic potential in each
material}. These are obtained from two independent standard bulk calculations
on the two materials.  The second term $\Delta V$, called the {\it lineup of
the average of the electrostatic potential} across the interface, accounts
for all the intrinsic interface effects. It is determined from a supercell
calculation with a model interface.  

However, the DFT-BOs suffer from two important limitations, namely, the
well-known DFT band-gap problem, and the use of an approximate functional to
model the exchange-correlation (XC) energy, such as the local density
approximation (LDA) or the generalized gradient approximation (GGA).  These
affect the value of the VBM and the CBM, and consequently the calculated BOs.
Orbital-dependent approaches~\cite{ Kummel2008} and many-body
perturbation-theory (MBPT) within the $GW$ approximation~\cite{ Hedin1965}
have been used to try to overcome these shortcomings~\cite{ Devynck2007,
Zhang1988, Zhu1991}. While the former may be less computationally demanding,
their reliability cannot be assessed a priori~\cite{ Kummel2008} in contrast
with MBPT. In Refs.~\cite{ Zhang1988, Zhu1991}, the lineup of the potential
$\Delta V$ was approximated by its DFT value, arguing that it depends
primarily on the charge distribution at the interface which is a ground state
quantity, hence little affected by many-body effects. Doing so, only the
band-structure contribution is modified:
\begin{equation}
\Delta E_{v,c}^{QP}=\Delta E_{v,c}^{DFT}+\Delta(\delta E_{v,c})
\label{eq:qpbo}
\end{equation}
where $\delta E_v$ ($\delta E_c$) is the quasiparticle (QP) correction at the
VBM (CBM): $\delta E_i=E_i^{QP}-E_i^{DFT}$ for $i=v,c$. It is important to
stress that these corrections, which are obtained from bulk calculations, are
the only additional ingredients that are required when DFT calculations of
the VBO and CBO already exist. 

Interestingly, for various semiconductor interfaces, the QP corrections of
the band edges are found to be almost the same on both sides~\cite{ Zhu1991}
leading to $\Delta(\delta E_v)\leq$ 0.2 eV in Eq.~(\ref{eq:qpbo}).  As a
result of this cancellation of errors, DFT is quite successful for the same
interfaces~\cite{ VandeWalle1987} with errors ranging from 0.1 to 0.5 eV,
despite its limitations mentioned above.  This relative success of DFT
explains why it has been widely used to predict the VBO for a wide range of
interfaces. And, when needed, the CBO was also predicted using a simple
scissor operator to correct the band gap to the experimental value. This
assumption was further motivated by the fact that MBPT calculations going
beyond $GW$ by including an approximate vertex correction ($GW\Gamma$) showed
that the valence-band-edge energy remained at its DFT value for silicon, the
whole correction going to the conduction bands~\cite{ DelSole1994,
Fleszar1997}.

However, when it comes to semiconductor-insulator or insulator-insulator
interfaces, it appears that the DFT errors on the VBO can be much more
important. For instance, for the Si/SiO$_2$ interface, the VBO are calculated
to be 2.3-3.3 eV~\cite{ Yamasaki2001, Watarai2004, Tuttle2004, Giustino2005a}
in noticeable disagreement with the experimental results of 4.3 eV~\cite{
Experiments}. It seems that, for such systems, the cancellation of errors is
not as good, emphasizing the need to go beyond DFT by including QP
corrections. The same is probably true for interfaces found in transistors or
organic photovoltaics.  Before applying such a highly-demanding method in a
predictive way, its quality needs to be assessed for an interface for which
the VBO and CBO are well-known experimentally, such as the Si/SiO$_2$ one.

In this Letter, the effect of QP corrections on the VBO and CBO at the
Si/SiO$_2$ interface is investigated. By performing a quasiparticle
self-consistent $GW$ (QS$GW$) calculation on a small interface model, we
demonstrate that the line up potential is well described in DFT. For two more
realistic models, the BOs calculated within DFT are corrected by computing
the QP corrections for bulk Si and SiO$_2$.  The latter are found to be
crucial to reach a good agreement with experimental results. Their effect can
not be accounted for by a simple scissor operator on top of DFT results.  The
inclusion of the vertex correction is shown not to affect the $GW$ results
significantly, whereas QS$GW$ does not improve the results.

Three different interface models are adopted in which ideal
$\beta$-cristobalite is matched to the Si(100) surface, taking the
theoretical value of bulk Si ($a_{Si}$=5.48 \AA) as the lattice constant.
The models I and II were generated in a previous work by two of us~\cite{
Giustino2005a} in order to calculate the BOs within DFT.  They consist of 11
Si monolayers and up to 10 SiO$_2$ molecular layers which guarantees well
converged BOs.  Model III is limited to 8 Si monolayers and 8 SiO$_2$
molecular layers in order to perform a QS$GW$ calculation which is quite
demanding computationally.  For bulk Si, the theoretical value $a_{Si}$ is
adopted for the lattice constant.  For SiO$_2$, we consider on the one hand
the cubic structure (space group $Fd\bar{3}m$) whose relaxed lattice constant
is $a_{SiO_2}$=7.43 \AA; and, on the other hand, a tetragonal structure
(space group $I4_1/amd$) in which two sides are strained to match the Si
lattice constant $a_{SiO_2}$=$\sqrt{2} a_{Si}$ as in the interface models of
Ref.~\cite{ Giustino2005a}, while the third is left free to relax reaching a
value of $c_{SiO_2}$=6.59 \AA. In both structures, the Si-O-Si bond angle is
180$^\circ$. Hereafter, the cubic and strained structures will be referred to
as c-SiO$_2$ and s-SiO$_2$, respectively.

All our calculations are performed using the ABINIT package~\cite{
Gonze2005}.  Only valence electrons are explicitly considered using
norm-conserving pseudopotentials \cite{ Troullier1991} to account for
core-valence interactions.  By coherence with the DFT calculations of the
BOs~\cite{ Giustino2005a}, the XC energy is also described within the
GGA~\cite{ Perdew1992}.

At the DFT level, the band gaps $E_g^{DFT}$ for the bulk systems are found to
be 0.7 eV for Si, 5.4 eV for c-SiO$_2$, and 5.1 eV for
s-SiO$_2$~\cite{note:bulk}.  For the interface models I and II, the VBOs
(CBOs) are calculated to be 2.6 and 2.5 eV (1.6 and 1.8 eV),
respectively~\cite{note:interface}.

Three approaches are used to compute the QP corrections.  In the first
approach ($GW$), the self-energy is calculated self-consistently by updating
the eigenenergies in both the dielectric matrices and the Green's functions
while keeping the DFT wavefunctions. In the second approach (QS$GW$), both
the eigenenergies and the wavefunctions are updated~\cite{Faleev2004}.  In
both approaches, the dielectric matrices are computed in the random-phase
approximation (RPA) using the sum over states formulation~\cite{
Adler1962-Wiser1963}. In the third approach, a vertex correction $\Gamma$ is
included in both the screening and the self-energy by going beyond the RPA
and including XC effects~\cite{ DelSole1994}, and only the eigenergies are
updated.  Details about the DFT-GGA kernel are provided in
Ref.~\cite{DalCorso1994}.  

Within $GW$, the frequency dependence of the dynamically screened Coulomb
potential $W$ is most often approximated using various plasmon pole models
(PPMs)~\cite{ Hybertsen1986, vonderLinden1988, Godby1989, Engel1993}.  The
advantage is not only to reduce the computational load, but also to obtain an
analytic expression for the self-energy. In fact, PPMs have proven to be very
effective in producing band gaps in good agreement with experiments.

However, while the QP correction to the gap $\delta E_g$ was found {\it not}
to be very sensitive to the choice of the PPM~\cite{ Arnaud2000}, we observe
that $\delta E_v$ and $\delta E_c$ {\it may vary from one PPM to another}, as
reported in Table~\ref{tab:ppm} for Si and c-SiO$_2$.  In particular, the
variation of $\delta E_g$ with the PPM is more pronounced in c-SiO$_2$ (up to
0.4 eV difference between the extreme cases) than in Si.  Since a precise
knowledge of the QP corrections at the band edges is required for the band
offsets calculations, {\it it is necessary to go beyond PPMs} taking
explicitly into account the frequency dependence of $W$.

\begin{table}[h]
\caption{Quasiparticle corrections (in eV) at the VBM ($\delta E_v$), at the
CBM ($\delta E_c$), and for the band gap ($\delta E_g$) for Si and c-SiO$_2$.
The corrections are calculated within $GW$ using the PPMs proposed by
Hybertsen and Louie (HL)~\cite{ Hybertsen1986}, von der Linden and Horsch
(vdLH)~\cite{ vonderLinden1988}, Godby and Needs (GN)~\cite{ Godby1989},
Engel and Farid (EF)~\cite{ Engel1993}, and without PPM.}
\label{tab:ppm}
\begin{center}
\begin{ruledtabular}
\begin{tabular}{llccccc}
 & & \multicolumn{1}{c}{HL} & \multicolumn{1}{c}{vdLH}
   & \multicolumn{1}{c}{GN} & \multicolumn{1}{c}{EF} &\multicolumn{1}{c}{no PPM}\\
\hline
Si        & $\delta E_v$ & $-0.6$  & $-0.6$  & $-0.4$ & $-0.6$ &$-0.4$ \\
          & $\delta E_c$ & $+0.1$  & $+0.1$  & $+0.2$ & $+0.1$ &$+0.2$ \\
          & $\delta E_g$ & $+0.7$  & $+0.7$  & $+0.6$ & $+0.7$ &$+0.6$ \\
\hline                                                          
c-SiO$_2$ & $\delta E_v$ & $-2.6$  & $-2.5$  & $-2.0$ & $-2.3$ &$-1.9$\\
          & $\delta E_c$ & $+1.3$  & $+1.1$  & $+1.5$ & $+1.2$ &$+1.5$\\
          & $\delta E_g$ & $+3.9$  & $+3.6$  & $+3.5$ & $+3.5$ &$+3.4$ \\
\end{tabular}
\end{ruledtabular}
\end{center}
\end{table}

In this work, the explicit frequency dependence is obtained using the
deformed contour integration technique~\cite{ Lebegue2003}.  The calculated
QP corrections are reported in Table~\ref{tab:qpcorr} for Si, c-SiO$_2$, and
s-SiO$_2$.

The QP gaps can easily be obtained from these results
$E_g^{QP}=E_g^{DFT}+\delta E_g$.  Within $GW$, $E_g^{QP}$=1.3 eV for Si and
8.8 eV for c-SiO$_2$. These results are in good agreement with the
experimental values of 1.2 and 8.9 eV~\cite{ DiStefano1971}, respectively.
While $GW\Gamma$ results (1.3 and 8.5 eV) are very close, the QS$GW$ results
(1.5 and 9.5 eV) are overestimated like systematically observed in
Ref.~\cite{ Shishkin2007}.

\begin{table}[h]
\caption{Quasiparticle corrections (in eV) at the VBM ($\delta E_v$), at the
CBM ($\delta E_v$), and for the band gap ($\delta E_g$)  for Si, c-SiO$_2$,
and s-SiO$_2$. The corrections are calculated within $GW$, $GW\Gamma$, and
QS$GW$.}
\begin{center}
\begin{ruledtabular}
\setlength{\tabcolsep}{1pt}
\begin{tabular}{clllclllclll}
&\multicolumn{3}{c}{Si} &
&\multicolumn{3}{c}{c-SiO$_2$} &
&\multicolumn{3}{c}{s-SiO$_2$} \\
\cline{2-4} \cline{6-8} \cline{10-12}
&\multicolumn{1}{c}{$_{GW}$} &\multicolumn{1}{c}{$_{GW\Gamma}$}
&\multicolumn{1}{c}{$_{\text{QS}GW}$} &
&\multicolumn{1}{c}{$_{GW}$} &\multicolumn{1}{c}{$_{GW\Gamma}$}
&\multicolumn{1}{c}{$_{\text{QS}GW}$} &
&\multicolumn{1}{c}{$_{GW}$} &\multicolumn{1}{c}{$_{GW\Gamma}$}
&\multicolumn{1}{c}{$_{\text{QS}GW}$} 
\\
\cline{2-4} \cline{6-8} \cline{10-12}
$\delta E_v$ & $-0.4$ & $+0.1$ &$-0.6$ && $-1.9$ & $-1.3$ & $-2.8$  && $-1.9$ & $-1.3$ &$-2.8$\\
$\delta E_c$ & $+0.2$ & $+0.7$ &$+0.2$ && $+1.5$ & $+1.8$ & $+1.3$  && $+1.4$ & $+1.8$&$+1.1$ \\
$\delta E_g$ & $+0.6$ & $+0.6$ &$+0.8$ && $+3.4$ & $+3.1$ & $+4.1$  && $+3.3$ & $+3.1$&$+3.9$ \\
\end{tabular}
\end{ruledtabular}
\end{center}
\label{tab:qpcorr}
\end{table}

Turning to $\delta E_v$ and $\delta E_c$, the comparison of our results with
previous calculations is not always straightforward. In most of the cases,
the focus is on the band gap and the QP corrections are not given explicitly
(the VBM is set to zero after correction). Besides, $\delta E_v$ and $\delta
E_c$ are very sensitive to the degree of convergence reached (the QP gap
converges much faster than the QP corrections), not to mention the effect of
the PPM (see Table~\ref{tab:ppm}).  For Si, our $GW$ and $GW\Gamma$ results
compare quite well with those of Fleszar and Hanke~\cite{ Fleszar1997} which
also do not rely on any PPM~\cite{ note:Fleszar}.  No such comparison can be
made for c-SiO$_2$. Note however that the QP corrections for c-SiO$_2$ and
s-SiO$_2$ differ by up to 0.2 eV, indicating a dependence of $\delta E_v$ and
$\delta E_c$ on the strain. This is at variance with the findings of
Ref.~\cite{ Zhu1989} in the case of Si under isotropic strain. A possible
explanation for this difference may be that the strain is not isotropic in
s-SiO$_2$~\cite{ note:bulk}.  Finally, note that the PPM proposed by Godby
and Needs~\cite{ Godby1989} leads to QP corrections (see Table~\ref{tab:ppm})
in excellent agreement with those of the contour deformation method, at
variance with the other PPMs. This finding might be generalized after proper
investigation.

While $GW$ and QS$GW$ lead to a lowering of the VBM of Si (slightly larger
for QS$GW$) compared to the DFT result, the inclusion of the vertex
correction brings it back to roughly its original value with a small shift of
0.1 eV, all the QP correction being on the conduction band. A similar result
was also found previously~\cite{ DelSole1994, Fleszar1997} giving some
motivation to the use of a scissor operator to compute the CBO within DFT.
For SiO$_2$, our results are very different. First, the VBM is also raised
when going from $GW$ to $GW\Gamma$, but it definitely does not reach the DFT
level back. This indicates the recovery of the DFT VBM with $GW\Gamma$ is a
coincidence in Si.  And, it definitely rules out the use of a simple scissor
operator for the computation of the BOs. Second, the lowering of the VBM in
QS$GW$ in much larger than in $GW$ (by about 0.9 eV). In fact, it seems that
the systematical overestimation of the band gap in QS$GW$ originates
essentially from a too strong lowering of the VBM. Note that the increase of
the CBM is slightly lower in QS$GW$ than in $GW$.

\begin{figure}[h]
\includegraphics{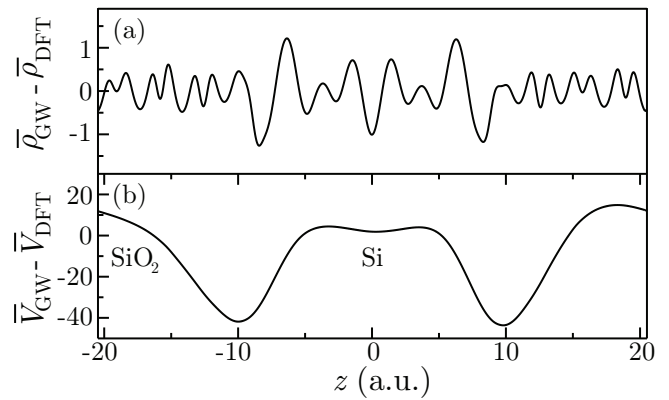}
\caption{Difference between DFT and QS$GW$ calculations for model III of (a)
the planar average of the electronic density and (b) the macroscopic average
of the local potential. The density is expressed in m$e$/a.u., and the
potential in meV.
\label{fig:rhov}}
\end{figure}

In order to compute the QP corrections to the BOs, the many-body effects on
$\Delta V$ also need to be investigated. This is done by comparing the
electronic density and the resulting $\Delta V$ calculated within DFT and
QS$GW$ for model III. In Fig.~\ref{fig:rhov}(a), the difference between the
planar average of the DFT electronic density $\bar{\rho}_{\mathrm{DFT}}(z)$
and the QS$GW$ one $\bar{\rho}_{\mathrm{GW}}(z)$ is reported. The QS$GW$
result differs only slightly from DFT, with a maximum change of about 1
m$e$/a.u. in the interface region.  The difference between the QS$GW$ and DFT
macroscopic average of the local potential
$\bar{V}_\mathrm{GW}-\bar{V}_\mathrm{DFT}$, which is reported in
Fig.~\ref{fig:rhov}(b), is about $\sim$ 45 meV in the interface region, and
less than 12 meV in the bulk regions. This gives rise to a net difference of
$\sim$ 20 meV in the lineup of the potential $\Delta V$.  Such a small
difference in $\Delta V$ suggests that the interfacial charge density and,
consequently, the associated dipole moments are well described within DFT.
This justifies the assumption that the line up potential can be taken to be
the same as in DFT.

Finally, using Eq.~(\ref{eq:qpbo}), the band offsets can be computed within
MBPT at the $GW$, $GW\Gamma$, and QS$GW$ levels. Our results are reported in
Table~\ref{tab:bo} and compared with the experimental ones.  Within $GW$ the
agreement is excellent for both the VBO and CBO (less than 0.3 eV
difference). The effect of the vertex correction is less than 0.1 eV on the
BOs. This results from a cancellation of the effects on each side of the
interface. At this stage, we cannot say whether this result can be
generalized to any interface.  In contrast, the effect of quasiparticle
self-consistency is more pronounced. Due to the large lowering of the VBM of
SiO$_2$, the VBO is increased by 0.7 eV compared to $GW$, leading to an
overestimation by up to 0.5 eV. At the same time, the CBO is slightly smaller
than in $GW$ leading to an underestimation by up to 0.6 eV. So that QS$GW$ is
the worst of the three approaches.

\begin{table}[h]
\caption{Quasiparticle band offsets (eV) for cubic and strained SiO$_2$ using
$GW$, $GW\Gamma$, and QS$GW$.}
\begin{center}
\begin{ruledtabular}
\begin{tabular}{cccccccccccccc}
    &        &      & & \multicolumn{3}{c}{cubic}
                    & & \multicolumn{3}{c}{strained}
                    & & \\
\cline{5-7} \cline{9-11}
    & $_{\text{Model}}$ & $_{\text{DFT}}$ & 
    & $_{GW}$ & $_{GW\Gamma}$ &$_{\text{QS}GW}$ &
    & $_{GW}$ & $_{GW\Gamma}$ &$_{\text{QS}GW}$& & $_{\text{Expt.}}$ \\
\hline
VBO &   I    & 2.6 & & 4.1  & 4.0&4.8        & & 4.1  & 4.0&4.8        & & 4.3   \\
    &   II   & 2.5 & & 4.0  & 3.9&4.7        & & 4.0  & 3.9&4.7        & &       \\
\hline
CBO &   I    & 1.6 & & 2.9  & 2.7&2.7        & & 2.8  & 2.7&2.5        & & 3.1   \\
    &   II   & 1.8 & & 3.1  & 2.9&2.9        & & 3.0  & 2.9&2.7        & &       \\
\end{tabular}
\end{ruledtabular}
\end{center}
\label{tab:bo}
\end{table}

In summary, we have investigated the band offsets at the Si/SiO$_2$ interface
using many-body perturbation theory. Starting from the BOs obtained within
DFT for two model interfaces, the QP corrections have been computed within
eigenenergy self-consistent $GW$ and $GW\Gamma$, and quasiparticle
self-consistent $GW$ taking the frequency dependence of the screened
potential explicitly into account (which is more reliable than plasmon pole
models).  The $GW$ corrections, which differ significantly from what can be
obtained using a scissor operator, produce BOs in excellent agreement with
experiment.  While the BOs obtained with and without vertex correction do not
differ significantly, those resulting from QS$GW$ present a larger deviation
from experiments.  These findings allow us to recommend the use of
eigenenergy self-consistent $GW$, which is less demanding.  Now that the
quality of the procedure has been assessed, it can be used in a predictive
way for heterojunctions of high technological interest such as those in
transistors or organic photovoltaics.

\begin{acknowledgments}
This work was supported by the EU FP6 and FP7 through the Nanoquanta NoE
(NMP4-CT-2004-50019) and the ETSF I3 e-Infrastructure (Grant Agreement
211956), and the project FRFC N$^\circ$. 2.4502.05.

\end{acknowledgments}

\bibliography{basename of .bib file}

\begin{thebibliography}{98}
\bibitem{note:bulk}
For Si, the gap is indirect with the VBM located at $\Gamma$ and the CBM
approximated by the $X$ point; while c-SiO$_2$ has a direct gap at $\Gamma$.
The lowering of the symmetry in s-SiO$_2$ results in a splitting of the
threefold-degenerate state at $\Gamma$ into two top valence states, the
highest corresponding to the $p_{z}$ orbital, and the other
(twofold-degenerate) to the $p_{x}$ and $p_{y}$ orbitals.  These two states
are separated by 0.27 eV, accounting basically for the difference in the band
gap between the two SiO$_2$ structures.
\bibitem{note:Fleszar}
The slight differences are probably due to the use of a different XC energy
functional at the DFT level (GGA in our case and LDA in the work of Fleszar
and Hanke).
\end{thebibliography}

\end{document}